\documentclass[twocolumn,prl,showpacs]{revtex4}
\usepackage{epsfig}
\usepackage{graphicx}
\usepackage{epsfig}
\usepackage{dcolumn}
\usepackage{bm}
\newcommand{\drv}{\textrm{d}}
\begin{document}

\title{Pattern formation and glassy phase in the $\phi^4$ theory with screened
electrostatic repulsion}
\author{Marco Tarzia$^{a}$ and Antonio Coniglio$^{a,b}$}
\affiliation{${}^a$ Dipartimento di Scienze Fisiche and INFN sezione di Napoli,
Universit\`{a} degli Studi di Napoli ``Federico II'',
Complesso Universitario di Monte Sant'Angelo, via Cinthia, 80126 Napoli, Italy}
\affiliation{${}^b$ Coherentia CNR-INFM}
\date{\today}

\begin{abstract}
We study analytically
the structural properties of a system with a short-range
attraction and a competing long-range screened repulsion. This model
contains the
essential features of the effective
interaction potential among charged colloids in
polymeric solutions and
provides novel insights on the equilibrium phase diagram of these systems.
Within the self-consistent Hartree approximation
and by using a replica approach,
we show that varying the parameters of the repulsive potential and the
temperature yields a phase coexistence, a lamellar and a glassy phase.
Our results
strongly
suggest that the cluster phase observed in charged colloids might
be the signature of an underlying equilibrium lamellar phase, hidden on 
experimental time scales.
\end{abstract}
\maketitle

{\em Introduction---}In the last few years there has been much interest in the
role of the inter-particle potential on controlling the structure and
the dynamics of colloidal 
suspensions~\cite{weitz,campbell,sator,sciortino,sciortino1} due to the
potential application of these systems
for designing new materials
with a wide range of viscoelastic properties. In charged colloidal
systems the effective interaction can be described in terms
of a short-range attraction and a long-range screened electrostatic
repulsion (well approximated by the DLVO potential \cite{dlvo}). The
competition between attractive and repulsive interactions on
different length scales stabilizes the formation of aggregates of an
optimal size ({\sl cluster phase}) characterized by a peak of the
structure factor around a typical wave vector, $k_m$,
that has been experimentally~\cite{weitz,campbell} and
numerically~\cite{sator,sciortino,sciortino1} observed. By an
appropriate tuning of the control parameters, the system
progressively evolves toward an arrested gel-like disordered state
({\sl colloidal gelation}). Although intensely studied both 
experimentally and numerically, a theoretical
understanding of the mechanisms underlying these phenomena
is still lacking and many gaps remain in our knowledge of the
equilibrium phase diagram of these systems.

In this letter we study analytically a $\phi^4$ model with
competition between a short-range attraction, described by the
Ginzburg-Landau Hamiltonian, and a long-range screened repulsion,
described by a Yukawa potential. Albeit schematically, this model
contains the essential features of the effective interaction among 
charged colloids and sheds new light on the structural properties of
these systems. Depending on the control parameters, there is a
region of the phase diagram where usual phase separation takes
place. Conversely, as the screening length and/or the strength of
the repulsion exceeds a threshold value, phase separation is
prevented. In this case, at moderately high temperature the
competition
between attraction and repulsion has the effect to produce modulated
structures, which in the terminology of particle systems, correspond
to the cluster phase. These modulated structures are the precursors
of a first order transition towards an equilibrium lamellar phase found
at lower temperatures.
By using a replica approach for systems without quenched
disorder~\cite{monasson}, and employing the Self-Consistent Screening
Approximation (SCSA)~\cite{bray}, we also show the presence of a
glass transition line in the low temperature region, once the first
order transition to the lamellar phase is avoided. Note that the
mechanism for the glass transition in this case is not due to the
presence of hard-core type of potential. Instead, it is due to the
formation of the modulated structures which order up to the size of
correlation length.
The geometric frustration, resulting in arranging such modulated
structures in a disordered fashion, leads to a complex free energy
landscape and, consequently, to a dynamical slowing down.

Our results suggest that the cluster phase observed in
colloidal suspensions should be followed, upon decreasing the
temperature (or increasing the volume fraction), by an equilibrium
periodic phase (a tubular or a lamellar phase, depending on the
volume fraction). If, instead, this ordered phase is avoided, a structural
arrest, corresponding to the gel phase observed in
the experiments and in numerical simulations,
should eventually occur. 
The existence of ordered phases in colloidal suspension and the fact that  
the transition to the gel phase could occur in a metastable liquid, are novel 
predictions, which have never been considered before.
Recently, motivated by our work,
de Candia {\em et al.}~\cite{tubi} have unambiguously 
shown the presence
of such ordered phases by using MD simulations on an atomistic model
system of charged colloids, interacting via the DLVO potential.

{\em Model and phase diagram---}We consider the standard three
dimensional $\phi^4$ field-theory with the addition of a repulsive
long-range Yukawa potential:
\begin{eqnarray} \label{eq:model}
\nonumber
&& \!\!\!\!\!\!\!\!\!\!\!\!
{\cal H} [\phi] = \int \drv^3 \mathbf{x} \left[ \frac{r_0}{2} \phi^2
(\mathbf{x}) + \frac{g}{4}
\phi^4 (\mathbf{x}) + \frac{1}{2}
(\nabla \phi (\mathbf{x}))^2\right] \\
&& \qquad +\, \frac{W}{2} \int \!\!\!\! \int \drv^3 \mathbf{x} \,
\drv^3 \mathbf{x'} \, \frac{e^{- | \mathbf{x} - \mathbf{x'}| / \lambda} \,
\phi(\mathbf{x}) \phi(\mathbf{x'})}
{|\mathbf{x} - \mathbf{x'}|},
\end{eqnarray}
where $\phi (\mathbf{x})$ is the scalar order parameter field. The
model has been also studied in~\cite{westhfal} in the context of
microemulsion. The parameters $W$ and $\lambda$ are, respectively,
the strength and the range of the repulsive potential. For $W=0$ we
obtain the canonical short-range ferromagnet. Interestingly, for
$\lambda \to \infty$ we recover the case of the Coulombic
interaction~\cite{emery,coniglio,kivelson,tarjus,wolynes,wolynes1,tarjus1}.
This model has been used to describe the phenomenology of a wide
variety of systems, where competing interactions on different length
scales stabilize pattern formations and the creation of spatial
inhomogeneities (for a review see~\cite{review}). These systems
include magnetic systems  and dipolar fluids characterized by
long-range Coulombic interactions~\cite{roland}, mixtures of block
copolymers~\cite{ohta}, water-oil-surfactant
mixtures~\cite{stillinger} and doped Mott insulator, including the
high $T_c$ superconductors~\cite{tranquada}. As a consequence, our
model allows to describe and to interpret in an unified fashion the
phenomenology of a wide variety of different systems.

\begin{figure}[ht] 
\begin{center}
\psfig{figure=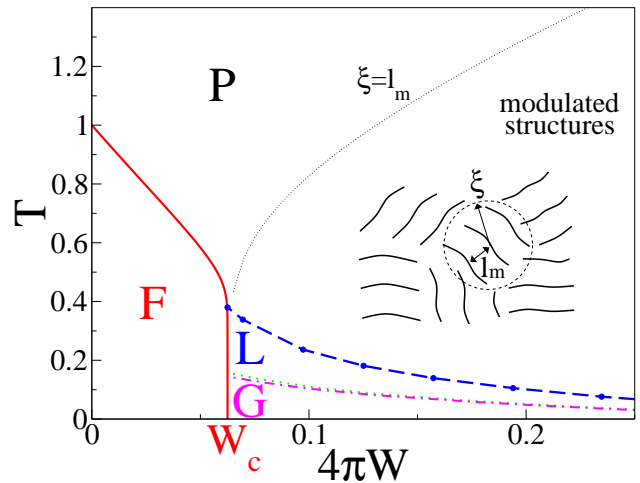,scale=0.31,angle=270}
\end{center}
\caption{(color online) Temperature ($T$)-repulsion
($4\pi W$) phase diagram 
of the model
for $\lambda = 2$ (and
$r_0=-1$), showing the relative positions of the paramagnetic (P),
ferromagnetic (F), lamellar (L) and glassy (G) phases. The value of
$g$ is such that 
$T_c (W=0) 
= 1$.   
The continuous (black) and the dashed (blue) curves, found
within the Hartree approximation, corresponds respectively to the
second-order phase transition from P to F, $T_c (W, \lambda)$, and
to the first-order transition from P to L, $T_L (W, \lambda)$. The
dotted (green) and the dashed-dotted (red) curves, found within the
SCSA, identify, the dynamical, $T_d(W, \lambda)$, and the ideal,
$T_K (W, \lambda)$, transition temperatures to G. The thin dotted (black)
line ($\xi = l_m$) marks a crossover temperature below which the
system establishes modulated structures, schematically sketched in
the figure.} \label{figure} 
\end{figure}

Here
we present only the main results, and we refer to a
paper in preparation for more details~\cite{forth}. In
Fig.~\ref{figure} the phase diagram 
of the model 
as function of the
temperature, $T$, and the strength of the repulsion, $W$, is
presented, for a fixed value of the screening length, $\lambda=2$.

{\em Ferromagnetic and lamellar phases---}We first solve the model
within the self-consistent Hartree approximation, which consists in
replacing the term $g\phi^4/4$ of Eq.~(\ref{eq:model}) with $3
\langle \phi^2 \rangle \phi^2 /2$ \cite{nota_NH}. 
This substitution allows
to compute the correlation function
$G(\mathbf{k}) = \langle \phi_{\bf k} \phi_{- {\bf k}} \rangle -
\langle \phi_{\bf k} \rangle \langle \phi_{- {\bf k}} \rangle$. 
In the
paramagnetic phase, $\langle \phi_{\bf k} \rangle = 0$, one obtains:
\begin{equation} \label{eq:gq}
T G^{-1} (\mathbf{k}) = r + k^2 + 4 \pi W \big / \left( \lambda^{-2} + k^2
\right),
\end{equation}
where the renormalized mass term, $r$, is defined as: $r \equiv r_0
+ 3 g \langle \phi^2 \rangle$. Since $\langle \phi^2 \rangle =
\int_{|\mathbf{k}|<\Lambda} \frac{\drv^3 \mathbf{k}}{(2 \pi)^3} \,
G(\mathbf{k})$, from Eq.~(\ref{eq:gq}) the following self-consistent
equation for $r$ is derived:
\begin{equation} \label{eq:hartree}
r = r_0 + 3 g \! \int_{|\mathbf{k}|<\Lambda} \!
\frac{\drv^3 \mathbf{k}}{(2 \pi)^3} \,
\frac{T}{r + k^2 + \frac{4 \pi W}{\lambda^{-2} + k^2}}
\end{equation}
($\Lambda$ is an ultraviolet cutoff). For convenience let us
define:
\begin{equation} \label{eq:wc}
4 \pi W_c  \equiv \lambda^{-4}.
\end{equation}
For  $W \le W_c$ and a fixed value of $\lambda$, we find a line of
ordinary second order critical points, $T_c(W,\lambda)$ (continuous
curve in Fig.~\ref{figure}),
separating a
high-temperature paramagnetic phase from  a low-temperature
ferromagnetic one. This transition is characterized by the
divergence of the susceptibility, i.e. $G^{-1}
(\mathbf{k}=\mathbf{0}) = 0$, with the usual Hartree critical
exponents (e.g., $\nu = 1$ and $\gamma=2$ in three dimensions).
Thus, for $W \le W_c$ usual phase separation occurs:
the effect
of the repulsive interaction is to decrease the value of $T_c$ from
the critical temperature of the standard Ginzburg-Landau model (for
$W=0$) to zero temperature (for $W \to W_c$).
Conversely,
above $W_c$
there is no phase separation. This result is quite important for
designing new materials as well as in
the experimental and numerical study
of colloidal systems, where it is crucial to distinguish the slowing
down due to colloidal gelation from that due to kinetic of phase
separation.
Interestingly, the threshold value $W_c$,
Eq.~(\ref{eq:wc}), coincides with that estimated for an atomistic
model system of charged colloids interacting via the DLVO
potential~\cite{sciortino1}.

In the 
limit, $\lambda \to
\infty$, we recover the already known results for the Coulombic case:
phase separation occurs only for
$W=0$~\cite{emery,coniglio,kivelson,tarjus,wolynes,wolynes1,tarjus1}.

For $W>W_c$, the model instead exhibits a first order transition
line $T_L(W,\lambda)$ (dashed curve of
Fig.~\ref{figure}), separating the paramagnetic phase from a
lamellar phase~\cite{brazovskii,tarjus1},
characterized by a spatially modulated order:
\begin{eqnarray}
\label{eq:kmax1} &&\langle \phi_{\mathbf{k}} \rangle = A \left(
\delta \left ( \mathbf{k} - \mathbf{k}_m \right) + \delta \left (
\mathbf{k} + \mathbf{k}_m \right) \right),
\end{eqnarray}
with amplitude $A$ and wave vector $k_m$, given by
\begin{eqnarray}
\label{eq:kmax}
&&k_m^2 = \sqrt{4 \pi} \left(  W^{1/2} - W_c^{1/2}
\right).
\end{eqnarray}
For $W>W_c$ the spinodal line of
the ``supercooled'' paramagnetic phase is located at $T=0$, where
the susceptibility diverges ($T G^{-1} (\mathbf{k}_m) = 0$).
As a result, the first-order transition to the lamellar phase can be
kinetically avoided and long-time glassy relaxations can be, instead,
observed~\cite{wolynes,wolynes1,tarjus1,tarjus,tarjus2,gonnella,reichman}.

{\em Glass transition---}In
order to analyze the glass transition in our model,
we employ a replica approach
formulated to deal with systems
without quenched disorder~\cite{monasson}.

The equilibrium free energy, $F = - T \ln Z$,
is relevant only if the system is able to explore ergodically
the entire phase space.
This is not the case, of course,
in the glassy phase, where the system is frozen
in metastable states.
In order to scan the locally stable field configurations,
we introduce an appropriate symmetry breaking field $\psi
(\mathbf{x})$, and compute the following partition function~\cite{monasson}:
\begin{equation}\label{pinning}
\! \tilde Z [\psi] \! = \! \int \!\! \mathcal{D} \phi \,
\exp
\left(
- \beta \mathcal{H} [\phi]
-\frac{u}{2} \! \int \! 
\drv^3 \mathbf{x} \left[ \psi (\mathbf{x}) - \phi (\mathbf{x})
\right]^2 \right),
\end{equation}
where $u>0$ denotes the strength of the coupling. The free energy
$\tilde f [\psi] = - T \ln \tilde Z [\psi]$
will be low if $\psi (\mathbf{x})$ equals to configurations
which locally minimize $\mathcal{H} [\phi]$.
Thus, in order to scan all metastable states we have to
sample all configurations
of the field $\psi$, weighted with $\exp(- \beta \tilde f [\psi])$:
\begin{equation}
\tilde F = \lim_{u \to 0^+}
\left[
\int \!\! \mathcal{D} \psi \,
\tilde f [\psi] \, e^{- \beta \tilde f [\psi]}
\bigg /
\int \!\! \mathcal{D} \psi \, e^{- \beta \tilde f [\psi] } \right]
\end{equation}
$\tilde F$
is a weighted average of the free energy in the various metastable states;
if there are only a few local minima, the limit behaves perturbatively
and $\tilde F$ equals the true free energy $F$.
However, in case of the emergence of an exponentially large number of
metastable states with large barriers between them, a nontrivial contribution
arises from the above integral even in the limit $u \to 0^+$ and $\tilde F$
differs from $F$. This allows to identify the complexity,
$\Sigma$, via the relation $F = \tilde F - T \Sigma$~\cite{monasson}.
In order to get an explicit expression for $\Sigma$ we introduce replicas:
\begin{equation}\label{eq:replica}
\tilde F (m) = - \lim_{u \to 0^+} \frac{T}{m} \ln \int \mathcal{D} \psi \,
\left( \tilde Z [\psi] \right)^m,
\end{equation}
from which, we obtain
$\tilde F = \partial m \tilde F (m) / \partial m |_{m=1}$ and
\begin{equation} \label{eq:sconf}
\Sigma = T^{-1} \left( \partial \tilde F(m) \big / \partial m \big|_{m=1} 
\right).
\end{equation}
Integrating Eq.~(\ref{eq:replica}) over $\psi$,
we get an action
which is formally equivalent to the (replicated) action
of a system in a quenched random field.
We can thus use the SCSA~\cite{bray}, a technique developed to
deal with such systems, which allows to
determine the correlators in the replica space.
The SCSA
amounts to introduce a $N$-component version of the
model and summing self-consistently
all the diagrams of order $1/N$~\cite{wolynes,wolynes1}.
Since the attractive coupling between replicas is symmetric with respect
to the replica index, one can assume
the following structure of the correlators in the replica space:
$G_{ab} (\mathbf{k}) = \left[ G (\mathbf{k}) - F
(\mathbf{k}) \right] \delta_{ab} + F (\mathbf{k})$,
i.e., with diagonal elements $G (\mathbf{k})$ and
off-diagonal elements $F (\mathbf{k})$.
For systems
with quenched disorder, this ansatz turns out to be equivalent to the
one-step replica symmetry breaking. While the diagonal correlator can be
interpreted as the usual one-time equilibrium correlation function,
$T G (\mathbf{k}) = \langle \phi_{\mathbf{k}} \phi_{\mathbf{-k}}
\rangle$, the off-diagonal term can be interpreted as measuring
the long-time correlations: $T F (\mathbf{k}) = \lim_{t \to
\infty} \langle \phi_{\mathbf{k}} (t) \phi_{\mathbf{-k}} (0) \rangle$.
Hence, $F (\mathbf{k})$ vanishes in the paramagnetic phase while is
finite in the glassy one.

The system undergoes a glass transition in the low temperature
region for $W>W_c$,
with exactly
the same nature of that found in mean-field models for glass-formers.
Lowering the temperature we first find a purely dynamical transition at
temperature $T_d$ (dotted curve in Fig.~\ref{figure}).
Here, the complexity, Eq.~(\ref{eq:sconf}), jumps
discontinuously from zero to a finite value, signaling the emergence of an
exponentially large number of metastable state.
The complexity decreases as the temperature is
decreased and vanishes at
$T_K$ (dashed-dotted curve in figure)
where the thermodynamical transition takes place.

\begin{figure}[ht]
\begin{center}
\psfig{figure=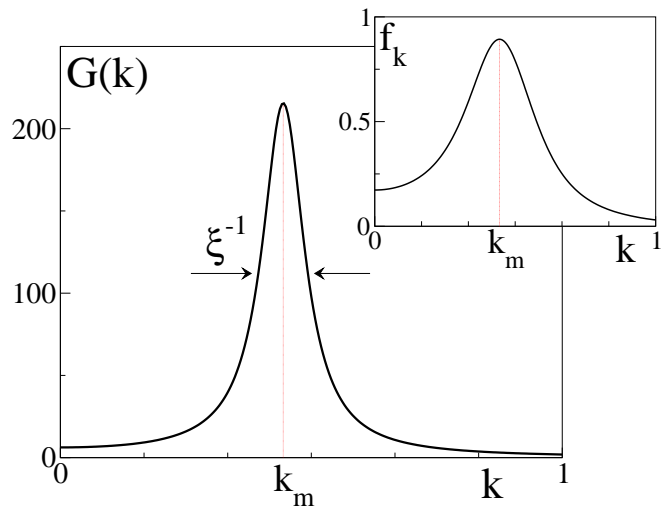,scale=0.31,angle=-90}
\end{center}
\caption{{\bf Main frame:} Momentum dependence of the
correlation function, $G(\mathbf{k})$,
for $4 \pi W = 0.2$ and $\lambda=2$ at $T=T_d$, showing
that it is peaked
around the typical modulation wave vector
$k_m$ with broadening $\xi^{-1}$, given by the inverse of the correlation
length. {\bf Inset:} Momentum dependence of the non ergodicity parameter
$f_{\mathbf{k}}$ for the same values of $W$, $\lambda$ and $T$.}
\label{fig:gk}
\end{figure}
{\em Structural properties---}The
correlation function, $G(\mathbf{k})$, is plotted in Fig.~\ref{fig:gk}
at the dynamical transition temperature $T_d$, showing a maximum at
$k_m$, defined in Eq.~(\ref{eq:kmax}), with width $\xi^{-1}$, given
by the inverse of the correlation length. The correlation function
in the real space reads: $G(|{\bf x}|) \sim e^{-|{\bf x}|/\xi} \,
\sin(k_m |{\bf x}|) / |{\bf x}|$. This expression implies that, although no
periodic order occurs ($\langle \phi_{\mathbf{k}_m} \rangle = 0$),
a lamellar structure of wave length $l_m = 2 \pi k_m^{-1}$ over a
finite range $\xi$ is formed (as sketched in Fig.~\ref{figure}). In
the glassy phase, $T \lesssim T_d$, one has that $\xi \gtrsim 2
l_m$~\cite{wolynes1}: thus these modulated structures form over a
length larger than their modulation length and become frozen. The
glass transition arises from the fact that there are many possible
configurations to arrange such modulated structures in a disordered
fashion, leading to a great number of metastable states. The
presence of this characteristic wave length dominates also the
dynamics, as indicated by the momentum dependence of the non
ergodicity parameter, $f_{\mathbf{k}} \equiv \lim_{t \to \infty}
\frac{\langle \phi_{\mathbf{k}} (t) \phi_{\mathbf{-k}} (0) \rangle}
{\langle \phi_{\mathbf{k}} (t) \phi_{\mathbf{-k}} (t) \rangle} =
\frac{F(\mathbf{k})}{G(\mathbf{k})}$, plotted in the inset of
Fig.~\ref{fig:gk} at $T_d$. The presence of a maximum at $k_m$
signals the fact that structural arrest is more pronounced over
length scales of order $l_m$. At higher temperatures, $T>T_d$, the
non ergodicity parameter, $f_{\mathbf{k}}$, vanishes and the glassy
phase disappears; correspondingly, $G(\mathbf{k})$ broadens and the
height of the peak decreases; hence, $\xi$ decreases until the
modulated structures fade continuously, approximately at a crossover
temperature where $\xi \approx l_m$ (dotted curve in
Fig.~\ref{figure}).

These results are intimately related to the phenomenology observed
in  
colloidal systems, where the competition between attraction and
repulsion leads to the formation of a phase of stable clusters at
low temperatures~\cite{weitz,campbell,sator,sciortino,sciortino1},
which is the analog of the modulated structures here found. Our
results suggest that the transition to the disordered gel phase,
numerically and experimentally observed in colloidal suspensions,
occurs, in fact, in a metastable liquid, due to the presence of an
underlying equilibrium lamellar phase (which might be more easily
detected by increasing the screening length $\lambda$). This novel
prediction has been confirmed by recent MD simulations of a model
systems of charged colloids in
$3d$~\cite{tubi};  
a clear indication of the presence of periodic phases was also
numerically found in $2d$~\cite{reatto}.


{\em Conclusions---}We have derived analytically the complete phase
diagram of a model with competition between short-range attraction
and long-range screened repulsion, which contains the essential
features of the interaction potential of charged colloids. To our
knowledge, this is the first theoretical investigations on these
systems. Our predictions have been confirmed by recent numerical
simulations and may be also experimentally checked.

We wish
to thank E. Del Gado,
A. Fierro, G. Gonnella and N. Sator. M.T. is indebted with G.
Tarjus for many useful discussions and remarks.
Work supported by  EU Network Number  MRTN-CT-2003-504712,
MIUR-PRIN 2004, MIUR-FIRB 2001, CRdC-AMRA, INFM-PCI.

\end{document}